\documentclass[10pt,letterpaper,twocolumn]{article} 

\usepackage{ol2}
\usepackage[draft]{hyperref}
\usepackage{amsmath}

\begin{document}

\twocolumn[ 

\title{Complete spatial characterization of an optical wavefront using a variable-separation pinhole pair}


\author{David T. Lloyd,$^{*}$ Kevin O'Keeffe, and Simon M. Hooker}

\address{
Department of Physics, University of Oxford, Clarendon Laboratory, \\ Parks Road, Oxford, OX1 3PU, UK
\\
$^*$Corresponding author: d.lloyd1@physics.ox.ac.uk
}

\begin{abstract}We present a technique for measuring the transverse spatial properties of an optical wavefront. Intensity and phase profiles are recovered by analysis of a series of interference patterns produced by the combination of a scanning `X-shaped' slit and a static horizontal slit; the spatial coherence may be found from the same data. We demonstrate the technique by characterizing high harmonic radiation generated in a gas cell, however the method could be extended to a wide variety of light sources.

This paper was published in Optics Letters and is made available as an electronic reprint with the permission of OSA. The paper can be found at the following URL on the OSA website: http://dx.doi.org/10.1364/OL.38.001173. Systematic or multiple reproduction or distribution to multiple locations via electronic or other means is prohibited and is subject to penalties under law.
\end{abstract}


 ] 

\noindent Characterizing the spatial and temporal properties of high harmonic generation (HHG) is crucial for understanding and controlling the nonlinear dynamics and quantum processes underlying this phenomenon. Indeed, it has been shown that macroscopic signatures in the spatial profiles of high harmonic radiation can reveal microscopic properties of matter under the influence of an intense laser field, such as the interference between different quantum trajectories\cite{Fingerprints}. Furthermore, many applications such as soft x-ray imaging\cite{CDICoh} and free electron laser seeding\cite{FELseed}, require detailed, spectrally resolved knowledge of the properties of high order harmonics.

Recently, a technique known as Spectral Wavefront Optical Reconstruction by Diffraction (SWORD) was described which is capable of retrieving frequency-resolved information on the spatial properties of HHG\cite{SWORD, SWORD2}. With this technique, a thin slit is scanned across the wavefront producing a series of diffraction patterns, the analysis of which allows the incident wavefront and transverse intensity profile of each harmonic order to be recovered. An alternative technique makes use of lateral shearing interferometry (LSI) for wavefront analysis, again providing the intensity and phase profiles of individual harmonics \cite{LSI}. Spectrally-resolved measurements of this type form a crucial step towards the complete space-time characterization and optimization of extreme ultraviolet (XUV) sources.

Although SWORD and LSI retrieve the phase and intensity profiles of each harmonic order, they provide no information on the spatial coherence of the source, an important parameter for applications such as coherent diffractive imaging\cite{Diff} and dynamic holography\cite{Holography}. Here we present a method which, to our knowledge, is the first capable of retrieving the transverse phase 
front, intensity profile and spatial coherence of spectrally resolved harmonics from a single scan. In keeping with the theme established by previous work, we dub our technique SCIMITAR: SCanning Interference Measurement for Integrated Transverse Analysis of Radiation.

\begin{figure}[htb]
\centerline{\includegraphics[width=7.5cm]{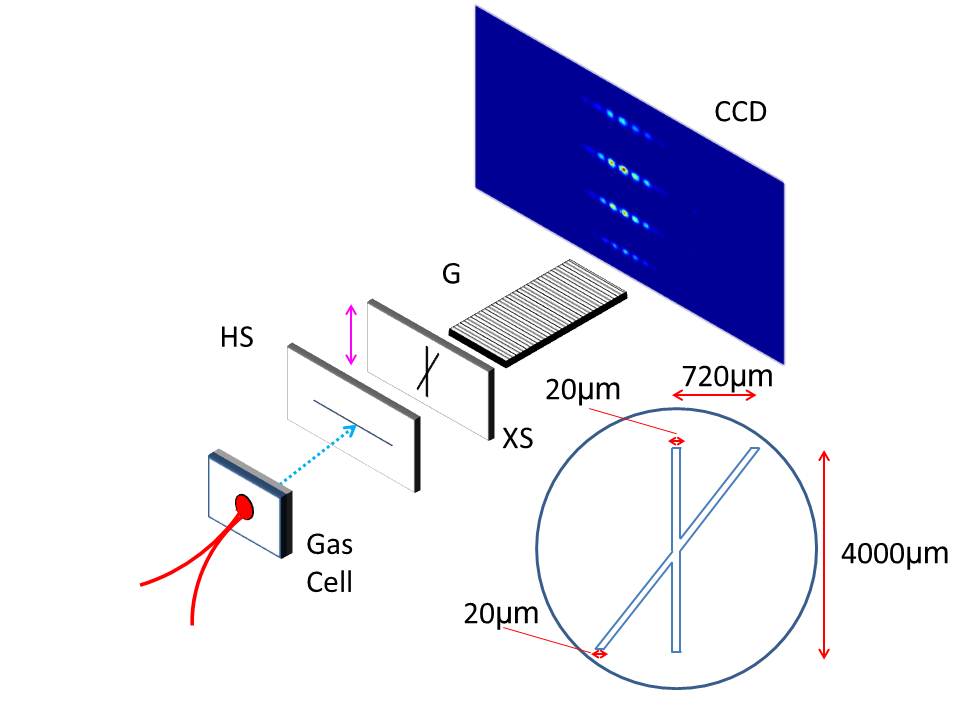}}
\caption{Experimental arrangement. High harmonics are generated in the gas cell by the focussed laser beam. They pass through the horizontal slit (HS), 20 $\mu$m by 4 mm in size, and the X-shaped slit (XS) located 5mm behind HS. A schematic of XS is shown in the lower right corner. In the case of the SWORD scan the HS and XS were replaced by a single 20$\mu$m diameter pinhole. The IR driving beam is blocked by a 400 nm thick aluminium foil located between the slits and grating (not shown). A spherical variable-line-spaced grating (G), diffracts the harmonic light, such that spectral information is encoded in the orthogonal direction to the spatial fringes on the x-ray CCD. The magenta arrow indicates the scanning direction of the X-slit.}
\label{EXP}
\end{figure}

As is the case with SWORD, SCIMITAR makes use of a diffracting slit for beam diagnosis. In our case we combine two apertures: a slit pair shaped like a tilted `X', as illustrated in Fig. \ref{EXP}, and a horizontal slit. If placed one after the other, with a separation small compared to the Rayleigh range of light diffracted from HS, they effectively form a pair of pinholes to the impinging beam. The pinhole separation is set by the vertical position of the X-slit. Scanning the X-slit in the vertical direction enables a continuous change in the pinhole separation in the horizontal direction. The X-slit is positioned so that one pinhole remains stationary at the beam centre during the scan. The beam properties of interest may be determined from the series of interference patterns produced. It should be noted that the accuracy with which the phase of the wavefront may be determined varies with the pinhole separation, and hence varies across the beam.

It can be shown that the interference pattern recorded by the illumination of a pinhole pair can be written as
\begin{eqnarray}
I(X)=I_0(X)\bigg[1+\mathcal{V}\cos\left(\frac{kXs}{z}-\frac{kx_0s}{z}+\Delta\phi\right)\bigg]
\label{IntDone}
\end{eqnarray}
where $I_0(X)$ is the envelope created by diffraction from the pinholes, $k$ is the wavevector of the incident light, $s$ the pinhole separation, $X$ is the horizontal position in the detector plane, $\Delta \phi$ is the phase difference between the portions of the wavefront sampled and $x_0$ is the distance between the edge of the detector and the centre of the pinhole pair. Eqn (\ref{IntDone}) takes into account the shift of the centre of the pinhole pair with respect to a static detector. The interference pattern may be analyzed by a well established Fourier transform technique [9] to yield the phase $\psi = \Delta \phi -kx_0 s/z$.  The total intensity of light at the two pinholes $I_1 + I_2$, where $I_1$ and $I_2$ are the respective beam intensities as measured at the pinhole locations,  is proportional to the value of the transform at zero spatial frequency The fringe visibility can be found from the relative magnitudes of the peaks in the Fourier transform associated with the diffraction envelope and the amplitude of the interference pattern. The transverse profiles of the HHG beam can therefore be retrieved by monitoring the variation of these quantities as a function of pinhole separation.

The spatial coherence of a beam may be measured using the Thomson-Wolf method\cite{ThompsonandWolf}: by recording, as a function of their separation, the visibility of the fringe pattern produced by a pair of pinholes. The visibilty of the interference pattern is related to the magnitude of the complex coherence factor (CCF), $\gamma$,  by\cite{BornandWolf}:
\begin{eqnarray}
\label{vis}
\mathcal{V}=\frac{I_{\rm{max}}(X)-I_{\rm{min}}(X)}{I_{\rm{max}}(X)+I_{\rm{min}}(X)}=\frac{2\sqrt{I_1}\sqrt{I_2}}{I_1+I_2}\gamma
\end{eqnarray}
where $I_{\mathrm{max}}$ and $I_{\mathrm{min}}$  are, respectively, the maxima and minima of the interference pattern.

To test these ideas, SCIMITAR was used to characterize the spatial properties of HHG from a gas cell using the arrangement shown schematically in Fig.\ \ref{EXP}. Laser pulses with an energy of 180 $\mu$J, duration 40 fs, of centre wavelength 800 nm, and repetition rate 1 kHz, were focused by a f/11.8 lens into a hollow nickel tube, backed by 170 mbar of argon. Odd harmonics of the laser wavelength up to order $q=29$ were generated. Spectra were recorded using a flat-field spectrograph consisting of a gold-coated flat-field grating with 1200 lines/mm and a cooled soft x-ray CCD camera (Princeton PIXIS-XO). A 45 second exposure time was used for each pinhole separation. In order to benchmark the transverse intensity and phase profiles recovered through SCIMITAR, a SWORD measurement was also performed, under the same experimental conditions, by transversely scanning a 20 $\mu$m diameter pinhole at the same longitudinal position as the X-slit.

\begin{figure}[htb]
\centerline{\includegraphics[width=7.5cm]{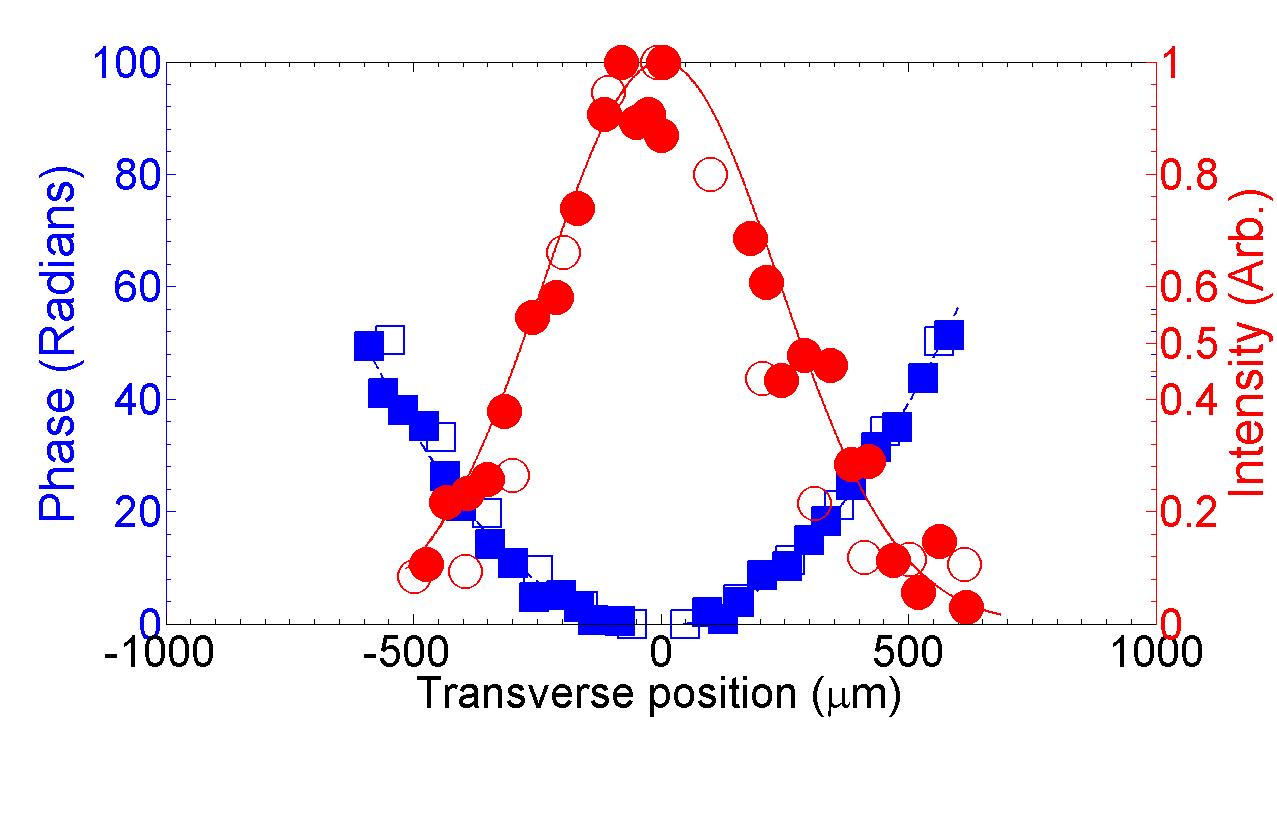}}
\caption{Recovered transverse intensity (circles) and phase (squares) profiles, for harmonic q=27, using SCIMITAR (closed symbols) and SWORD (open symbols) technique. Gaussian and parabolic fits to, respectively, the intensity and phase profiles are shown for the SCIMITAR data. Errorbars are smaller than the symbol size for both intensity and phase measurements.}
\label{Int27}
\end{figure}

To illustrate our results we plot in Fig. \ref{Int27} the intensity and phase profiles of harmonic order 27 obtained by SCIMITAR and SWORD techniques. It may be seen that excellent agreement is obtained between the measured SCIMITAR and SWORD data. The transverse variation of the wavefront phase measured by SCIMITAR is well approximated by a parabola, indicating that the $q=27$ harmonic has a diverging wavefront of radius 613 mm; this is similar to the distance between the X-slit and gas cell, which was measured to be 620$\pm$2 mm. A summary of fits to SCIMITAR profiles for other harmonic orders is presented in Table 1. 

It should be noted that the phase is obtained using SCIMITAR data in a different manner than that for SWORD. In the  latter, the transverse gradient of the wavefront phase is related to the position of the centre of the diffraction pattern from the pinhole; the wavefront phase must then be found by integrating the gradient across the beam. However, SCIMITAR is different in that the positions of the interference peaks are determined by the phase difference $\Delta \phi$ between the two pinholes. Since one of the pinholes remains in a fixed position during the scan, SCIMITAR gives directly  the phase $\psi$. In principle the linear term $(kx_0s/z)$ can be removed by careful measurement of $x_0$ and $z$. However, for the purposes of this paper we have removed all linear contributions to $\psi$ from the recovered phase.

\begin{table}[tb]
\centering
\begin{tabular}{|l|c|c|r|}
\hline
Harmonic & Intensity & Radius of & Coherence \\ 
Order & FWHM & Curvature & length\\
 &($\mu$m)& (mm)&  ($\mu$m)\\ \hline
23 & 525 & 599 & 760 \\ \hline
25 & 465 & 601 & 642\\ \hline
27 & 425 & 613 & 630 \\ \hline
29 & 418 & 634 & 566\\ 
\hline
\end{tabular}
\caption{Summary of the results of fits  to the transverse variations of the intensity, phase, and spatial coherence obtained from SCIMITAR measurements. The coherence length is defined as the full width half maximum of a gaussian fit to a plot of $\gamma$ versus $s$.}
\end {table}

\begin{figure}[htb]
\centerline{\includegraphics[width=7.5cm]{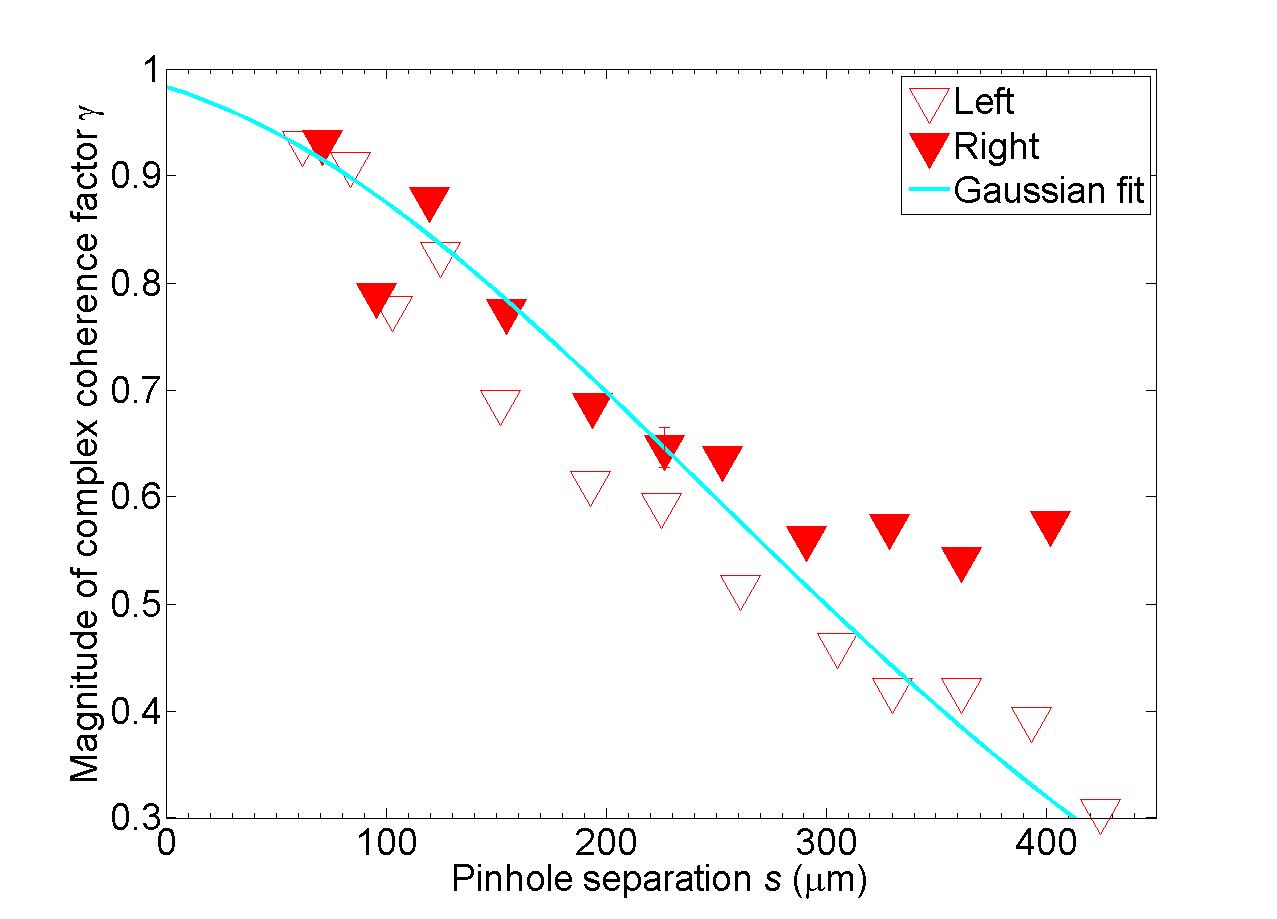}}
\caption{Variation of $\gamma$ with pinhole separation for $q=27$, deduced from the SCIMITAR measurements. A typical error bar is shown, found from the standard error of four repeat measurements. A Gaussian fit to all data points displayed is shown by the solid cyan line.}
\label{Vis27a}
\end{figure}
For each harmonic order $q$ the fringe visibility was determined as a function of the pinhole separation $s$; when combined with the measured transverse intensity profile, Eqn (\ref{vis}) yielded $\gamma$. The variation of $\gamma$ with $s$ is shown in Fig. \ref{Vis27a} for $q=27$.  The CCF shown here represents a lower limit limit since fluctuations of the beam pointing during the measurement would also reduce the fringe visibility by a factor  which increases with the pinhole separation. The data shown in Fig. \ref{Vis27a} exhibits a slight left-right asymmetry; this may arise from asymmetry in the driving laser beam or transverse variations of the gas density in the cell. The observation of this subtle effect illustrates the sensitivity of the SCIMITAR method. To our knowledge, the possibility and implications of an asymmetry in spatial coherence have not been considered previously.

In accordance with the van Cittert-Zernike theorem\cite{BornandWolf}, the line of best fit in Fig. \ref{Vis27a} corresponds to the CCF variation of an incoherent source, with a Gaussian intensity distribution of diameter (FWHM) 24$\mu$m, located at the gas cell.

For all harmonic orders measured, the widths of the Gaussian fits to the spatial coherence are broader than those of the transverse intensity profiles, showing that the harmonics are highly coherent, as reported previously\cite{Ditmire1}. It may be seen that with increasing harmonic order the beam size decreases, the radius of curvature increases and the coherence length decreases.

Finally we note that previous studies of the spatial coherence of HHG \cite{Ditmire1, Bartels}, employed a relatively small number of slit-pairs placed in the beam in turn, and as a consequence the transverse coherence was measured at only a few points. In contrast, SCIMITAR allows the coherence variation across the wavefront to be measured in much finer detail. 

We have presented and demonstrated a novel means of characterizing the transverse spatial properties of light: SCIMITAR. The simultaneous detailed measurement of the transverse variation of the  intensity, phase, and coherence  possible with SCIMITAR will allow better characterization of many types of radiation used to image and probe matter.

We are grateful for financial support from the Engineering and Physical Sciences Research
Council under grant EP/G067694/1.


\end{document}